\documentclass[a4paper,11pt]{article}
\pdfoutput=1
\usepackage{jheppub}
\usepackage{subfig}
\usepackage{amsmath,amssymb,bm,graphicx,epsf,colordvi}
\usepackage{slashed}
\def\bea#1\eea{\begin{align}#1\end{align}} 
\usepackage{diagbox}
\usepackage{mathtools}
\usepackage{lipsum}
\usepackage{soul}
\usepackage{enumitem}
\usepackage{braket}
\usepackage{bm}
\allowdisplaybreaks 
\usepackage{color}
\usepackage[dvipsnames]{xcolor}

\usepackage{mathrsfs}
\usepackage{appendix}
\usepackage{bm}
\usepackage{lipsum}

\newcommand{\bef}{\begin{figure}[hbt]\centering}
\newcommand{\eef}{\end{figure}}
\usepackage{mathtools}
\usepackage{booktabs}
\usepackage{graphicx}
\graphicspath{ {./figure/} }
\usepackage{caption}
\usepackage{lipsum}

\newcommand{\beq}{\begin{equation}}
\newcommand{\eeq}{\end{equation}}
\def\bea#1\eea{\begin{align}#1\end{align}}

\newcommand{\f}{\frac}
\def \be  {\begin{equation}}
\def \ee  {\end{equation}}
\def \ba  {\begin{eqnarray}}
\def \ea  {\end{eqnarray}}

\newcommand{\nn}{\nonumber}

\allowdisplaybreaks

\usepackage{graphicx}

\def\Fig#1{Fig.~{\ref{#1}}}
\DeclareRobustCommand{\Sec}[1]{Sec.~\ref{#1}}

\DeclareRobustCommand{\eq}[1]{Eq.~(\ref{#1})}

\definecolor{darkpurple}{rgb}{0.5, 0.2, 0.8}
\definecolor{or}{rgb}{0.88,0.43,0.02}
\definecolor{darkgreen}{rgb}{0.13,0.55,0.13}

\usepackage{stackengine}

\preprint{MIT-CTP 5788}
\title{Revisiting Single Inclusive Jet Production:\\ Small-$R$ Resummation at Next-to-Leading Logarithm}
\author[a]{Kyle Lee,}
\affiliation[a]{Center for Theoretical Physics, Massachusetts Institute of Technology, Cambridge, MA 02139}

\author[b]{Ian Moult,}
\affiliation[b]{Department of Physics, Yale University, New Haven, CT 06511}

\author[c]{Xiaoyuan Zhang}
\affiliation[c]{Department of Physics, Harvard University, Cambridge, MA 02138}

\emailAdd{kylel@mit.edu,ian.moult@yale.edu,xiaoyuanzhang@g.harvard.edu}

\abstract{The precision description of jet production plays an important role in many aspects of collider physics. In a recent paper we have presented a new factorization theorem for inclusive small radius jet production. The jet function appearing in our factorization theorem exhibits a non-standard renormalization group evolution, which, starting at next-to-leading logarithm (NLL), differs from previous results in the literature. In this paper we perform a first phenomenological study using our newly developed formalism, applying it to compute the spectrum of small radius jets in $e^+e^-\to J+X$ at NLL.  We compare our results with previous predictions, highlighting the numerical impact of previously neglected terms throughout phase space.  Our approach can be used for a variety of different collider systems, in particular, $ep$ and $pp$ collisions, with broad applications to the jet substructure program. Most importantly, since our factorization theorem is valid to all orders, the approach developed here will enable NNLL resummation of small radius logarithms in inclusive jet production, extending the precision of jet substructure calculations.
}

\begin{document}

\maketitle

\section{Introduction}

Precision predictions for processes involving high-p$_T$ jets and their substructure play a crucial role in collider experiments, with applications ranging from searches for beyond the Standard Model physics, to improving the understanding of nucleon structure. Arguably the simplest process is the inclusive production of a single high-$p_T$ jet, identified with a clustering algorithm, most often anti-$k_T$ \cite{Cacciari:2008gp}. This process has been computed to high perturbative orders \cite{Currie:2018xkj,Czakon:2021ohs,Czakon:2019tmo,Goyal:2024tmo,Goyal:2023zdi,Bonino:2024wgg,Bonino:2024qbh}, and is also a building block for jet substructure observables measured on inclusive jets \cite{Hannesdottir:2022rsl,Lee:2023xzv,Craft:2022kdo,Chien:2018lmv,Lee:2022ige,Lee:2023npz,Aschenauer:2019uex,Kang:2018qra,Cal:2020flh,Kang:2020xyq,Kang:2018vgn,Kang:2018jwa,Kang:2019prh,Cal:2021fla,Cal:2019gxa,Kang:2016ehg,Cal:2019hjc,Kang:2017mda,Mehtar-Tani:2024smp}. 

Due to the large Lorentz boosts achieved in collisions at the LHC, jets with small radius parameters, $R$ are used in many phenomenological applications. Measurements exist for values of $R$ as small as $R=0.1$ \cite{ALICE:2019qyj,CMS:2020caw}, and jets with radius $R=0.2$ are commonly used in heavy ion studies to reduce backgrounds (see e.g. \cite{ALICE:2023waz,CMS:2021vui}). For small $R$ jets, one can capture the leading behavior in the small-$R$ limit through the use of factorization theorems, where the inclusive jet process factorizes into a jet function, describing the infrared measurement associated with the clustering algorithm, along with any jet substructure measurements, and a hard function, which is process dependent, but independent of the jet structure. This factorization is much in analogy with the factorization theorems for inclusive hadron production \cite{Collins:1981ta,Bodwin:1984hc,Collins:1985ue,Collins:1988ig,Collins:1989gx,Collins:2011zzd,Nayak:2005rt}, where one has a standard factorization into an inclusive hard function, and a universal fragmentation function. However, in the case of a jet identified with an infrared and collinear safe jet algorithm, the infrared dynamics can be computed in perturbation theory. This factorization approach has two advantages: First, it simplifies the required calculations since the jet functions are universal and can be combined with any hard function.  Second, it allows a resummation of logarithms associated with the jet radius, allowing for improved theoretical predictions \cite{Dasgupta:2014yra,Dasgupta:2016bnd,Kang:2016mcy,Kang:2016ehg,Liu:2018ktv}. There have been two primary ways of resumming these jet radius logarithms, one using generating functions/parton showers \cite{Dasgupta:2014yra,Dasgupta:2016bnd,vanBeekveld:2024jnx,vanBeekveld:2024qxs}, and another using factorization theorem derived~\cite{Kang:2016mcy,Kang:2016ehg} with effective field theory~\cite{Bauer:2000ew,Bauer:2000yr,Bauer:2001ct,Bauer:2001yt,Rothstein:2016bsq}. In general, the advantage of an all orders factorization theorem is that, if correct, it can be easily extended to higher perturbative orders.

In a recent paper~\cite{Lee:2024icn}, we have shown that the factorization formula presented in \cite{Kang:2016mcy} neglected certain terms, which arise starting at NLL. These terms arise due to a distinction between jets identified with a jet algorithm (at least those of the $k_T$ family), and single particle hadron states appearing in fragmentation. This issue was also identified and studied in \cite{vanBeekveld:2024jnx,vanBeekveld:2024qxs}. After identifying this issue, we then derived an all-order factorization theorem that correctly incorporates these additional terms, while maintaining the universality of the hard function. We showed that while the renormalization group evolution satisfied by the jet function is not the standard Dokshitzer-Gribov-Lipatov-Altarelli-Parisi (DGLAP) equation governing the evolution of fragmentation functions, it is still entirely determined by the timelike splitting functions. Therefore, while the equation itself is more complicated, it does not require the calculation of new anomalous dimensions. Our factorization theorem was inspired by a similar factorization theorem for the (projected) energy correlators \cite{Dixon:2019uzg,Chen:2020vvp}, where explicit perturbative calculations exist at higher perturbative orders \cite{Belitsky:2013ofa,Dixon:2018qgp,Luo:2019nig,Henn:2019gkr}, allowing for verification of the non-trivial renormalization group structure. We further carried out confirmation of our factorization by explicitly computing the anomalous dimensions with two-loop jet function calculations.

In this paper, we show how to apply our factorization theorem in practice and solve the associated renormalization group equations, opening the door for phenomenological applications. To illustrate our new factorization theorem in the simplest possible setting, we compute the inclusive jet energy spectrum in $e^+e^-$ collisions, $e^+e^-\to J+X$, at NLL. While this calculation has appeared previously~\cite{Neill:2021std,Kang:2016mcy}, and been compared with LEP data in~\cite{Chen:2021uws}, it was performed using the framework of \cite{Kang:2016mcy,Kang:2016ehg}, which we have shown to be incomplete. We numerically explore the impact of the neglected terms throughout phase space, finding that they grow towards lower jet energies and give moderate corrections. Most importantly, we believe that our calculation resolves numerous issues in the literature at NLL, setting the stage for calculations at NNLL.

An outline of this paper is as follows. In \Sec{sec:review} we provide a brief review of our new factorization theorem for inclusive jet production derived in Ref.~\cite{Lee:2024icn}, as well as the associated renormalization group evolution for the jet function. In \Sec{sec:solution} we show how to iteratively solve the renormalization group evolution equations for the jet function. In \Sec{sec:pheno} we apply our factorization theorem to make phenomenological predictions for $e^+e^-\to J+X$, and highlight the impact of neglected terms throughout phase space. We conclude in \Sec{sec:conc}.

\section{Factorization Theorem for Small-$R$ Jet Production}\label{sec:review}

We begin by briefly reviewing the factorization theorem for small radius jet production derived in Ref.~\cite{Lee:2024icn}.  We consider the inclusive production of a high-$p_T$ jet, identified with the anti-$k_T$ jet algorithm \cite{Cacciari:2008gp} (or any algorithm from the $k_T$ family). 
While our factorization theorem applies for generic initial states ($pp$, $ep$, $ee$) and, the jet function as well as its renormalization group evolution are universal, we will focus on the specific case of $e^+e^-$ collisions for simplicity.
We will also use the case of $e^+e^-$ collisions as a phenomenological example in \Sec{sec:pheno}. 
See~\cite{Lee:2024icn} for an explanation on how the factorization can be generalized to high-$p_T$ jet production.

Consider the process $e^+e^-\to J+X$, with center of mass energy $Q$. To leading power in the jet radius parameter $R$, the jet spectrum differential in $z_J=2E_J/Q$ is described by the factorized expression
\begin{align}
\label{eq:HJ}
\frac{d\sigma}{dz_J} =&\int dx\,dz~ \vec{H}\left(x,\frac{Q^2}{\mu^2},\mu \right) \cdot \vec{J}\left(z,\ln\frac{x^2 Q^2 R^2}{4\mu^2},\mu \right) \delta(z_J - x z)\\
=&\int_{z_J}^1 \frac{dx}{x}~ \vec{H}\left(x,\frac{Q^2}{\mu^2},\mu \right) \cdot \vec{J}\left(\frac{z_J}{x},\ln\frac{x^2 Q^2 R^2}{4\mu^2},\mu \right)\,.\nn
\end{align}
Here $\vec H$ is the inclusive hard function vector in flavor space, identical to the one in the factorization theorem for single identified hadron fragmentation \cite{Collins:1981ta,Bodwin:1984hc,Collins:1985ue,Collins:1988ig,Collins:1989gx,Collins:2011zzd,Nayak:2005rt}. The hard function satisfies the timelike DGLAP equation \cite{Gribov:1972ri,Dokshitzer:1977sg,Altarelli:1977zs}
\begin{align}
  \label{eq:hardRG}
  \frac{d \vec{H} (x, \frac{Q^2}{\mu^2},\mu)}{d \ln \mu^2} = - \int_x^1 \frac{dy}{y} \widehat P_T(y,\mu) \cdot \vec{H}\left(\frac{x}{y}, \frac{Q^2}{\mu^2},\mu\right) \,,
\end{align}
governed by the timelike splitting kernels
\begin{align}
  \label{eq:splitK}
  \widehat P_T = 
  \begin{pmatrix}
    P_{qq} &\hspace{0.15cm}  P_{qg}
\\
    P_{gq} &\hspace{0.15cm} P_{gg}
  \end{pmatrix} \,,
\end{align}
which are known to three-loops \cite{Mitov:2006ic,Mitov:2006wy,Chen:2020uvt}. The moments of the timelike splitting kernels define the timelike DGLAP anomalous dimensions $\gamma_T(N)$~\cite{Mitov:2006wy,Mitov:2006ic,Moch:2007tx,Almasy:2011eq}
\begin{equation}
\label{eq:gammaPTrelation}
\gamma_T(N)\ \equiv\ - \int_0^1 dy \, y^{N-1} \, \widehat{P}_T(y).
\end{equation}

The jet function appearing in \eq{eq:HJ} describes the infrared dynamics associated with the jet algorithm. For an infrared and collinear safe algorithm, such as anti-$k_T$, it can be computed in perturbation theory. Although it will not be crucial for this paper, the jet function can be given an operator equation using the formalism of SCET \cite{Bauer:2000ew,Bauer:2000yr,Bauer:2001ct,Bauer:2001yt,Rothstein:2016bsq}. It is defined for quark and gluon jets as
\begin{align}
J_q\left(z=p_J^- / \omega, \omega_J, \mu\right)&=\frac{z}{2 N_c} \operatorname{Tr}\left[\frac{\slashed{\bar{n}}}{2}\left\langle 0\left|\delta(\omega-\bar{n} \cdot \mathcal{P}) \chi_n(0)\right| J X\right\rangle\left\langle J X\left|\bar{\chi}_n(0)\right| 0\right\rangle\right]\,, \\
J_g\left(z=p_J^- / \omega, \omega_J, \mu\right)&=-\frac{z \omega}{2\left(N_c^2-1\right)}\left\langle 0\left|\delta(\omega-\bar{n} \cdot \mathcal{P}) \mathcal{B}_{n \perp \mu}(0)\right| J X\right\rangle\left\langle J X\left|\mathcal{B}_{n \perp}^\mu(0)\right| 0\right\rangle\,.\nn
\end{align}
Here $\chi_n$ and $\mathcal{B}_{n \perp \mu}$ are gauge invariant quark and gluon fields in SCET, $\mathcal{P}$ is the momentum operator, and $J$ denotes the jet state, as identified with a specific jet algorithm.

The key new feature of the factorization theorem in \eq{eq:HJ}, is the non-trivial convolution structure between the hard and jet functions. This modification, as compared to the standard Mellin convolution structure for hadron fragmentation, is due to the modification of the IR measurement. Crucially, due to this modified structure, while the hard function obeys the standard DGLAP evolution, the jet function \emph{does not}. However, the renormalization group evolution of the jet function is fixed by renormalization group consistency, and takes the form
\begin{align}
  \label{eq:jetRG}
 \frac{d \vec{J}\left(z,\ln\frac{Q^2 R^2}{4\mu^2},\mu \right)}{d \ln \mu^2} = \int_z^1 \frac{dy}{y}  \vec{J}\left(\frac{z}{y},\ln\frac{y^2 Q^2 R^2}{4\mu^2},\mu \right) \cdot \widehat P_T(y)\,.
\end{align}
We therefore see a key advantage of this factorization theorem, namely it remains fixed by the standard timelike DGLAP kernels, although in a non-standard form. Since the DGLAP splitting kernels are known to high orders, this allows the resummation of jet radius logarithms to be extended to higher perturbative orders.

When transformed to moment space, \eq{eq:jetRG} implies that the moments of the jet function
\begin{align}
\label{eq:hadDGmom}
\vec{J} \left(N,\ln\frac{Q^2 R^2}{4\mu^2},\mu\right) &\equiv \int_0^1 dz\, z^{N}\,\vec{J} \left(z,\ln\frac{Q^2 R^2}{4\mu^2},\mu\right)\,,
\end{align}
obey the evolution equation
\begin{align}
  \label{eq:incljetRG}
\frac{d \vec{J}\left(N,\ln\frac{Q^2R^2}{4\mu^2}, \mu\right) }{d \ln \mu^2} = \int_0^1 dy\, y^{N} \vec{J} \left(N,\ln\frac{y^2 Q^2R^2}{4\mu^2}, \mu\right) \cdot \widehat P_T(y,\mu) \,, 
\end{align}
which is equivalent to that for the projected energy correlators \cite{Dixon:2019uzg,Chen:2020vvp}. See Ref.~\cite{Lee:2024icn} for extensive discussion on the connection between inclusive jet production and the projected energy correlators.

\section{Jet Function Renormalization Group Equation at NLL}\label{sec:solution}

Due to the ubiquity of the DGLAP evolution equations, there exist many programs for their efficient solution (e.g. \cite{Salam:2008qg,Botje:2010ay,Bertone:2013vaa}), which are used in numerous phenomenological applications. Higher-order analytic solutions have also been explored \cite{Bolzoni:2013rsa,Bolzoni:2012ii}. On the other hand, the evolution equation in \eq{eq:jetRG} is less standard. While its solutions for fixed moments have been explored in the context of the (projected) energy correlators \cite{Dixon:2019uzg,Chen:2020vvp,Chen:2023zlx,Jaarsma:2023ell}, the complete equation in the momentum fraction space has not previously been studied. Due to its appearance in the description of single inclusive jet production, we find it important to develop techniques for its solution.

For the typical jet radii used in phenomenology at current colliders, we find that it is sufficient to solve the RG equation, by iterating it to high loop orders. We study the numerical convergence in $e^+e^-\to J+X$ in \Sec{sec:pheno}, and find that this approach converges well. We can produce high loop iterative solutions to the RG equation, by solving the inclusive jet renormalization group equation in \eq{eq:incljetRG}, and then performing an inverse Mellin transform. This gives rise to the following form for the
renormalized jet function
\begin{align}
\label{eq:Jren}
J_i\left(z,\ln\frac{Q^2 R^2}{4\mu^2},\mu \right) =& \sum_{n=0}^\infty \sum_{m=0}^n \frac{a_s^n\, L^m}{m!}J_i^{(n,m)} = \delta(1-z) + a_s \left[J_i^{(1,0)} + \textcolor{blue}{J_i^{(1,1)}}L\right] \nn\\
&\hspace{-3cm}+ a_s^2 \left[J_i^{(2,0)} + \textcolor{blue}{J_i^{(2,1)}}L + \textcolor{blue}{J_i^{(2,2)}}\frac{L^2}{2}\right]\nn\\
&\hspace{-3cm}+a_s^3 \left[J_{i}^{(3,0)}+\textcolor{blue}{J_{i}^{(3,1)}}L + \textcolor{blue}{J_{i}^{(3,2)}}\frac{L^2}{2}+ \textcolor{blue}{J_{i}^{(3,3)}}\frac{L^3}{3!}\right] \nn\\
&\hspace{-3cm}+a_s^4 \left[J_{i}^{(4,0)}+\textcolor{blue}{J_{i}^{(4,1)}}L + \textcolor{blue}{J_{i}^{(4,2)}}\frac{L^2}{2}+ \textcolor{blue}{J_{i}^{(4,3)}}\frac{L^3}{3!}+ \textcolor{blue}{J_{i}^{(4,4)}}\frac{L^4}{4!}\right] \nn\\
&\hspace{-3cm}+a_s^5 \left[J_{i}^{(5,0)}+\textcolor{blue}{J_{i}^{(5,1)}}L + \textcolor{blue}{J_{i}^{(5,2)}}\frac{L^2}{2}+ \textcolor{blue}{J_{i}^{(5,3)}}\frac{L^3}{3!}+ \textcolor{blue}{J_{i}^{(5,4)}}\frac{L^4}{4!}+ \textcolor{blue}{J_{i}^{(5,5)}}\frac{L^5}{5!}\right] + \mathcal{O}(a_s^6)\,,
\end{align}
to $5$-loop order, where $L = \ln 4\mu^2/(Q^2R^2)$ and $a_s = \alpha_s/(4\pi)$. Note that each $J_i^{(n,m)}$ is a function of momentum fraction $z$ only. It is straightforward to generalize this to higher loop order and the terms in blue are predicted by RG evolution. To N$^k$LL accuracy, terms of the form $J_i^{(n,m)}$ with $n\geq m \geq n-k$ are predicted by the inclusive jet RG for $m\geq 1$.

To NLL, their iterative structure is given as
\begin{align}
\label{eq:NLLiter}
J_i^{(m,m)}(z) =\,& P_{ki}^{(0)} \otimes J_i^{(m-1,m-1)} + (m-1)\beta_0 J_i^{(m-1,m-1)}\,, \hspace{2cm}  \text{ for } m\geq 1\\
J_i^{(m,m-1)}(z) =\,& P_{ki}^{(1)}\otimes J_k^{(m-2,m-2)}  + P_{ki}^{(0)}\otimes J_k^{(m-1,m-2)} \hspace{2.75cm}\text{ for } m\geq 2\nn\\
&\hspace{-1cm}+ (m-1)\beta_0 J_{i}^{(m-1,m-2)}+ (m-2)\beta_1 J_{i}^{(m-2,m-2)}- 2P_{ki}^{(0)}\otimes (J_{k}^{(m-1,m-1)}\textcolor{red}{\ln y})\,,    \nn
\end{align}
where $\otimes$ represents the standard Mellin convolution. We highlight the $\textcolor{red}{\ln y}$ in the last term of $J_i^{(m,m-1)}(z)$, where $y$ represents the convolution variable. This term is precisely the consequence of modified RG structure for inclusive jet and dropping it corresponds to the standard DGLAP RG predictions. Taking moments of this convolution structure shows sensitivity to the derivative terms of the timelike anomalous dimensions as explained in Ref.~\cite{Lee:2024icn}. For higher resummation orders, higher derivative terms also exist -- for example, ones involving $\ln^2 y$ at NNLL. The boundary functions are computed from the explicit fixed-order calculation. For $k_T$-type algorithm, they are given as~\cite{Kang:2016mcy,Neill:2021std}
\begin{align}
\label{eq:NLOJet}
J_q^{(0,0)}(z) =&J_g^{(0,0)}(z)= \delta(1-z)\,,\nn\\
J_q^{(1,0)}(z) =& -2\ln z\,\left[P^{(0)}_{qq}(z)+P^{(0)}_{gq}(z)\right]
 \nn \\&
 -\Bigg\{2C_F\left[2(1+z^2)\left(\f{\ln(1-z)}{1-z}\right)_++(1-z)\right]
 \nn\\&
-\delta(1-z) 2C_F\left(\f{13}{2}-\f{2\pi^2}{3}\right)+2 P^{(0)}_{gq}(z) \ln(1-z)+2C_F z\Bigg\}\,,\nn\\
J_g^{(1,0)}(z) =& -2\ln z\,\left[P^{(0)}_{gg}(z)+2 N_f P^{(0)}_{qg}(z)\right]
\nn \\&
-\Bigg\{\frac{8 C_{A}\left(1-z+z^{2}\right)^{2}}{z}\left(\frac{\ln (1-z)}{1-z}\right)_{+}-2\delta(1-z)\bigg(C_{A}\left(\frac{67}{9}-\frac{2 \pi^{2}}{3}\right)
\nn\\&
-T_{F} N_{f}\left(\frac{23}{9}\right)\bigg)+4 N_{f}\left(P^{(0)}_{q g}(z) \ln (1-z)+2T_{F} z(1-z)\right)\Bigg\}\,,
\end{align}
where the leading timelike splitting functions $P_{ij}^{(0)}$ from $P_{ij} =\sum a_s^{n+1} P_{ij}^{(n)}$ are given as
\begin{align}
\label{eq:LOP}
P^{(0)}_{q q}(z)&=2C_F\left[\frac{1+z^2}{(1-z)_{+}}+\frac{3}{2} \delta(1-z)\right]\,, \nn\\
P^{(0)}_{g q}(z)&=2C_F \frac{1+(1-z)^2}{z}\,,\nn\\
P^{(0)}_{g g}(z)&=2 C_A\left[\frac{z}{(1-z)_{+}}+\frac{1-z}{z}+z(1-z)\right]+\frac{\beta_0}{2} \delta(1-z)\,,\nn\\
P^{(0)}_{q g}(z)&=T_F\left[z^2+(1-z)^2\right]\,.
\end{align}
The $n$-loop jet functions $J_{i}^{(n)}$ can be expressed as harmonic polylogarithms \cite{Remiddi:1999ew} and zeta values of weight up to $2n$, for which there exist numerous fast numerical implementations \cite{Vollinga:2004sn,Gehrmann:2001pz,Bauer:2000cp}. Simplified expressions for the coefficients for both quark and gluon jets are provided with the ancillary material. Since the jet functions are universal, we expect that these will be useful in a number of phenomenological applications. This approach to solving the RG equations can be applied in an identical manner for the case of identified hadron production in jets, or in combination with the measurement of a jet substructure observable.

\section{Phenomenological Predictions for $e^+e^-\to J+X$}\label{sec:pheno}

To illustrate our factorization theorem and renormalization group evolution in a phenomenological process, we consider the simplest case, $e^+e^-\to J+X$. This process is also interesting since the inclusive jet spectrum has recently been measured on archival LEP data~\cite{Chen:2021uws}. However, we emphasize that our factorization theorem applies more generally, and it would be interesting to apply it also to the case of $ep$ and $pp$ collisions. To achieve NLL accuracy, we need the two-loop timelike splitting functions, and the one-loop inclusive hard functions for $e^+e^-$. These are well known, and are summarized for example in \cite{Mitov:2006ic,Mitov:2006wy}.

\begin{figure}[t]
    \begin{center}
\includegraphics[scale=0.245]{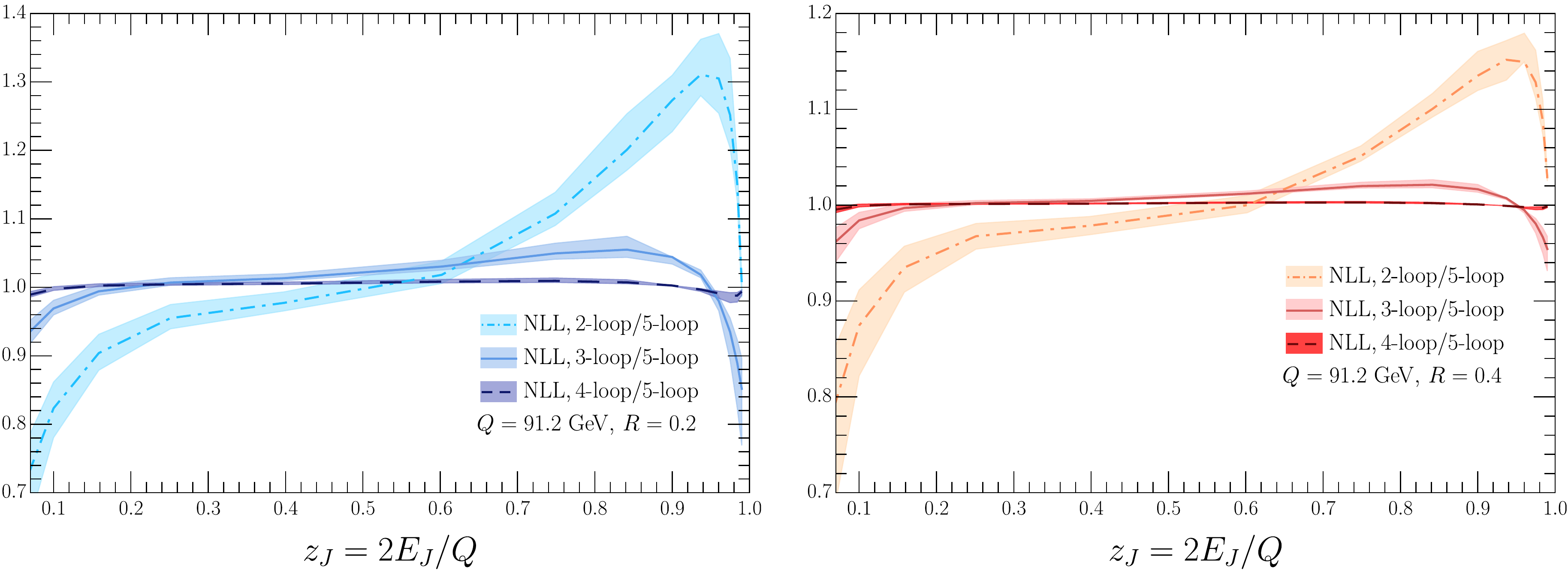}
     \caption{The convergence of NLL inclusive jet spectrum in $e^+e^-\to J+X$ with different orders in the iterative solution ($R=0.2$ in the left panel and $R=0.4$ in the right panel). For most of the jet energy fraction region, going from 4-loop to 5-loop has a subpercent difference.}
     \label{fig:convergence}
     \end{center}
 \end{figure}
 
Since the goal of this paper is to highlight the resummation structure of our new factorization theorem, we do not focus on producing complete phenomenological results, which would require the incorporation of non-perturbative effects and threshold resummation~\cite{Dasgupta:2007wa,Lee:2024esz,Neill:2021std}. Instead, we simply wish to highlight the structure of our perturbative result for $e^+e^-\to J+X$ with small-$R$ resummation at NLL, and its comparison to previous results in the literature. Throughout this section we will use two values of the jet radius, $R=0.2$ and $R=0.4$ to highlight the effects of jet radius resummation. Measurements with wide ranges of $R$ have been performed at the LHC, see e.g.  \cite{ALICE:2019qyj,CMS:2020caw}, and small values of $R$ are often used in jet substructure studies in heavy ion collisions. We therefore believe that our results in this simplified context of $e^+e^-$ collisions are representative of their phenomenological impact more generally.

We first begin by studying the convergence of our iterative solution. In \Fig{fig:convergence} we compare the inclusive jet spectrum in $e^+e^-\to J+X$ at NLL for $R=0.2$ and $R=0.4$ at LEP energy $\sqrt{s}=Q=91.2$ GeV~\cite{Chen:2021uws}, computed by taking the iterative solution of the jet function given in~\eq{eq:NLLiter} at 2, 3, 4, and 5 loops to evolve between the jet scale $\mu_J\sim QR/2$ to the hard scale $\mu_H\sim Q$ and then convolve with the hard function according to~\eq{eq:HJ}. To compare, we take the ratio of lower order solution to 5-loop solution. The scale uncertainties are determined by independently varying the jet and hard scales by a factor of 2 around their canonical values, while assuming the same scale choice for individual curves for different loop solutions. We observe excellent convergence for both $R=0.2$ and $R=0.4$, where the ratio between the 4-loop and 5-loop predictions have a subpercent deviation for most of the jet energy fraction and only around $\sim1\%$ level difference in the threshold or small-$z_J$ region. As expected from the size of the enhanced size of the small-R logarithm, we find a slower convergence for $R=0.2$, relative to $R=0.4$, though our fast convergence illustrates that $5$-loop solution is sufficient for phenomenological applications.

 \begin{figure}
    \begin{center}
 \includegraphics[scale=0.25]{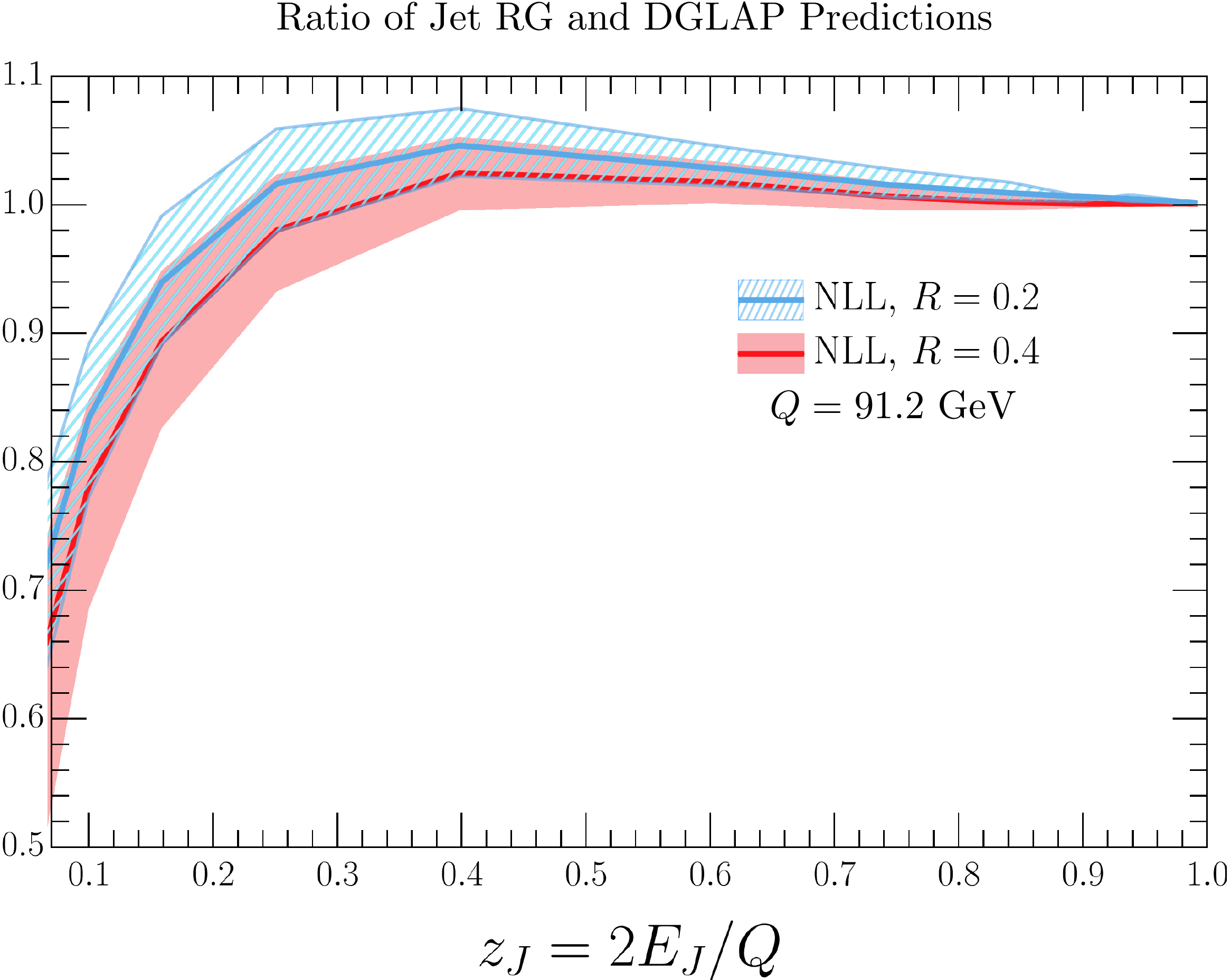}
     \caption{The ratio of the inclusive jet spectrum in $e^+e^-\to J+X$ computed using the factorization theorem presented in this paper, and previous results using the DGLAP convolution, at NLL, for both $R=0.2$ and $R=0.4$. The difference between the two approaches clearly increases at smaller jet energy fractions. This effect becomes more pronounced as the jet radius decreases, due to the enhanced size of the jet radius logarithms.}
     \label{fig:ratio}
     \end{center}
 \end{figure}

In \Fig{fig:ratio}, we compare the spectrum using our new inclusive jet factorization theorem with the one using the standard timelike DGLAP factorization for the fragmentation process. 
The ratio plot shows that the two approaches differ significantly at small values of the jet energy fraction, where the difference between the two convolution structures are accentuated. When approaching the threshold limit, i.e. $z_J\to 1$, expanding the convolution variable around $1$ becomes a good approximation and the difference between the two factorization formulas disappears.  This can also be seen by comparing the anomalous dimension for the moments of the jet function in the two approaches. These were computed in \cite{Lee:2024icn}, and it was shown that they converge for large moments, or equivalently, in the threshold limit. We also find that these features are enhanced for smaller $R=0.2$ relative to the $R=0.4$ case as expected. While we believe that these effects are already interesting for NLL phenomenology, they are crucial for getting an NLL baseline under control to extend the accuracy of the calculation to NNLL.

\begin{figure}
    \begin{center}
\includegraphics[scale=0.25]{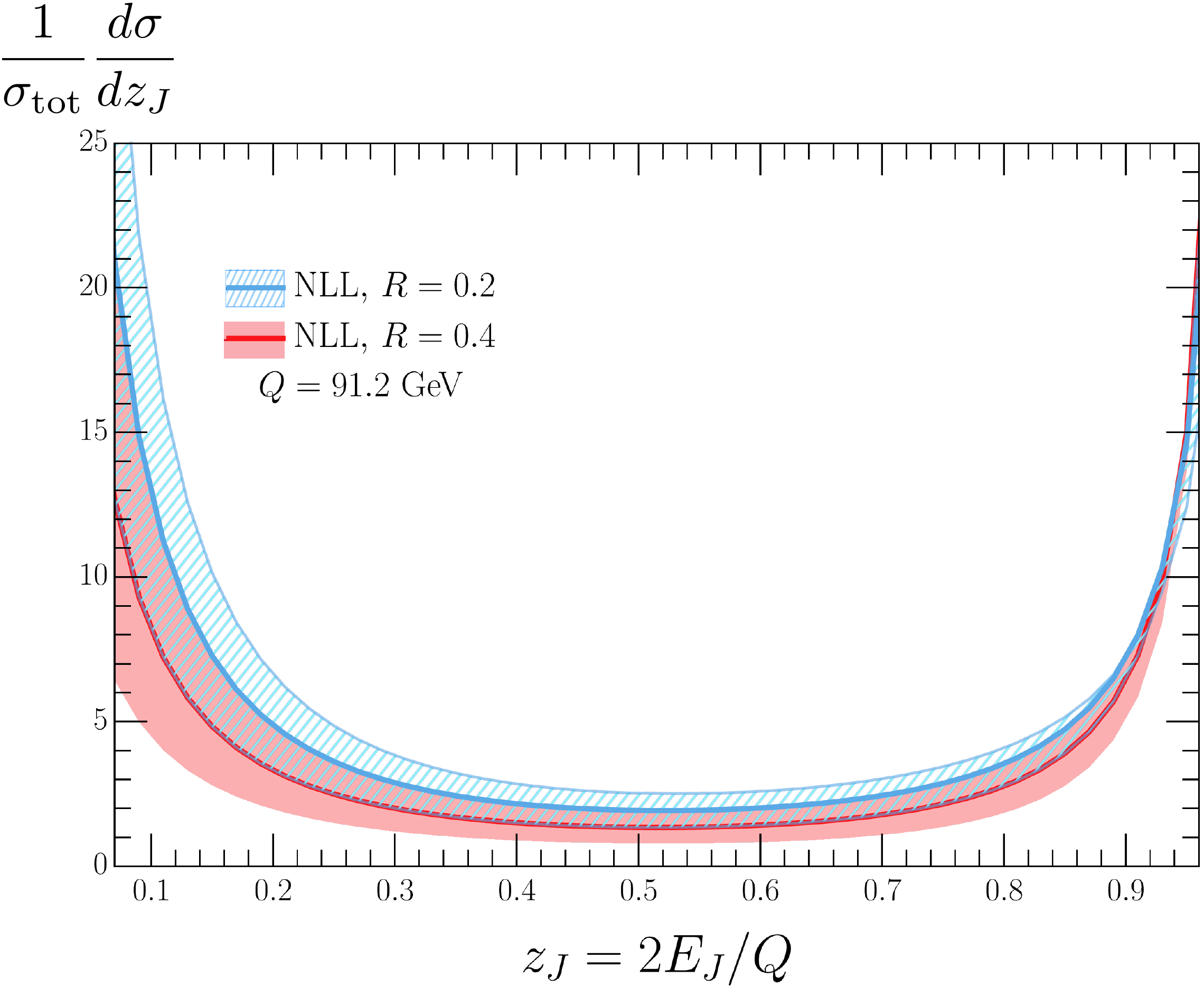}
     \caption{The inclusive jet spectrum in $e^+e^- \to J + X$ with NLL small-$R$ resummation computed using the new factorization theorem presented in this paper for both $R = 0.2$ and $R = 0.4$. The uncertainty bands are the envelope of hard variations and jet variations by a factor of 2. }
     \label{fig:compare}
     \end{center}
 \end{figure}

Finally in \Fig{fig:compare}, we present predictions for the inclusive jet spectrum in $e^+e^- \to J + X$ at NLL for $R = 0.2$ and $R = 0.4$ at LEP energy. The distribution is normalized to the partonic total cross-section $\sigma_{\rm{tot}}=1+3a_s C_F+\mathcal{O}(a_s^2)$. The relative enhancement observed for $R = 0.2$ compared to the $R = 0.4$ case can be attributed to the enhanced logarithmic terms. It will be important to perform a complete phenomenological analysis using our new framework by including non-perturbative effects and threshold resummation, to compare with the data of Ref.~\cite{Chen:2021uws}. It would also be interesting to perform a detailed comparison to the parton shower approach of Ref.~\cite{vanBeekveld:2024qxs}. We leave both these directions to future work.

\section{Conclusions}\label{sec:conc}

Inclusive jet production is a fundamental process in collider physics, and its precise description is an important component of the jet substructure program. In a recent paper~\cite{Lee:2024icn} we have presented a new factorization theorem describing inclusive small radius jet production, which allows for the resummation of small radius logarithms. As compared to previous approaches, where a standard DGLAP evolution was assumed for the jet functions, we showed that beyond LL, the standard DGLAP evolution is modified giving rise to a new evolution equation. 

In this paper we showed how to solve this equation iteratively at any specific logarithmic accuracy, opening the door to its phenomenological applications. Our evolution equation is identical to that for the moments of the $N$-point projected energy correlators, and building on techniques used for their study, we solved this equation iteratively.  As a first illustration of our formalism to phenomenology, we applied it to compute the inclusive small radius jet spectrum in $e^+e^-\to J+X$. While this process is of intrinsic interest as a precision test of QCD, particularly in light of recent re-analyses of LEP data~\cite{Chen:2021uws}, we view this primarily as an illustration of our formalism in the simplest possible context of $e^+e^-$ collisions. We compared our results to previous predictions for this process, finding sizable differences, particularly at smaller values of the jet energy fraction. This is important both for NLL phenomenology, but more importantly, it is crucial to have a correct NLL baseline to be able to proceed to higher perturbative orders. 

Although we have focused on the case of $e^+e^-$ colliders in this paper, the only modification required for other colliders is the hard function. It would therefore be interesting to apply our formalism to phenomenology in $ep$ and $pp$ collisions. The hard functions are known analytically for both $e^+e^-$ \cite{Mitov:2006wy} and $ep$
\cite{Bonino:2024wgg,Bonino:2024qbh,Goyal:2024tmo}, making these simple phenomenological targets for applying our formalism.

The results of this paper are a stepping stone towards extending the description of inclusive jet production, and jet substructure more generally, to NNLL. We believe that our new factorization theorem clarifies a number of previous issues in the literature. With this formalism in place, we can now move on to the extension to NNLL. A key advantage of our formalism is that the renormalization group equations are entirely determined by the DGLAP splitting functions, which are known to the required perturbative order for NNLL resummation \cite{Mitov:2006ic,Chen:2020uvt}. The remaining required perturbative ingredients are the two-loop constants for the inclusive jet function. These will be presented in a future paper.  We believe that our results lay a firm foundation for improving the perturbative description of jets in collider physics.

\acknowledgments
We thank Hannah Bossi, Terry Generet, Yibei Li, Felix Ringer, Rene Poncelet, and Nobuo Sato for useful discussions. K.L. was supported by the U.S. Department of Energy, Office of Science, Office of Nuclear Physics from DE-SC0011090. I.M. was supported by the DOE Early Career Award DE-SC0025581. XY.Z. was supported in part by the U.S. Department of Energy under contract DE-SC0013607.

\appendix

\section{Perturbative Ingredients}\label{sec:pert}

In this Appendix, we summarize the perturbative ingredients required for the NLL calculation of the inclusive jet spectrum in $e^+e^-$. While all these ingredients are well known in the literature, we summarize them here with common conventions, with the hope that it will be useful for future investigations. Calculations were performed with the help of the \texttt{HPL} \cite{Maitre:2005uu},  \texttt{HarmonicSums} \cite{Ablinger:2009ovq}, and \texttt{MT} \cite{Hoschele:2013pvt} packages.

We expand the splitting functions as
\begin{align}
P_{ij} =\sum_{n=0}^{\infty} a_s^{(n+1)} P_{ij}^{(n)}\,.
\end{align}
The LO timelike splitting functions were given in~\eq{eq:LOP}. The NLO splitting functions are standard, and relatively lengthy, so we do not reproduce them here. They can be found in \cite{Mitov:2006wy,Mitov:2006ic}. The NLO jet functions were also given already in the main text in \eq{eq:NLOJet}.

\subsection{Hard Functions}\label{sec:hard}
To achieve NLL resummation, we need the NLO hard function. This is the same hard function from the hadron fragmentation process \cite{Mitov:2006ic}, which depends on the parton flavor and parton energy fraction $x=\frac{2p\cdot q}{Q^2}$,
where $q$ is the total momentum and $p$ is the parton momentum.
We expand the hard function as
\begin{align}
H_i\left(x,\frac{Q^2}{\mu^2},\mu\right) =\sum_{n=0}^{\infty}  a_s^n H_i^{(n)}=\sum_{n=0}^{\infty} \sum_{m=0}^{n} \frac{a_s^n\ln^m \frac{\mu^2}{Q^2}}{m!} H_i^{(n,m)}\,,
\end{align}
where $i=q,g$ denotes the partonic channel and the second equality highlights the logarithmic structure of this expansion. Similar to the jet function, the log-enhanced terms are predicted by the RG structure, and thus we only present the constants, i.e. $H_i^{(n,0)},\, n\geq 0$.  For $e^+e^-\to q\bar q$ process, the leading order hard function is simply $H_i^{(0,0)} (x) = \{2\delta(1-x), 0\}$. At one-loop, we have
\begin{align}
\frac{1}{2}  H_{q}^{(1,0)}(x) = &\  \frac{\alpha_s}{4 \pi} C_F 
\Bigg[
\left(\frac{4 \pi ^2}{3}-9\right) \delta(1-x) +4 \left[\frac{\ln(1-x)}{1-x} \right]_+
\nn\\
&\
+\left(4 \ln
   (x)-\frac{3}{2}\right)\left(2 \left[\frac{1}{1-x}\right]_+-x-1\right)-\frac{9 x}{2}-2 (x+1) \ln
   (1-x)+\frac{7}{2}
\Bigg] \,,
\nn\\
H_{g}^{(1,0)}(x) = &\, \frac{\alpha_s}{4 \pi}C_F \Bigg[
\frac{4 \left(x^2-2 x+2\right) \ln (1-x)}{x}+\frac{8 \left(x^2-2 x+2\right) \ln   (x)}{x}
\Bigg] \,.
\end{align}
The factor $1/2$ in front of the quark channel indicates for identical contribution from anti-quark, since we do not distinguish quark and anti-quark flavor.

\bibliography{EEC_ref.bib}

\providecommand{\href}[2]{#2}\begingroup\raggedright\begin{thebibliography}{10}

\bibitem{Cacciari:2008gp}
M.~Cacciari, G.~P. Salam, and G.~Soyez, {\it {The anti-$k_t$ jet clustering
  algorithm}},  {\em JHEP} {\bf 04} (2008) 063,
  [\href{http://arxiv.org/abs/0802.1189}{{\tt arXiv:0802.1189}}].

\bibitem{Currie:2018xkj}
J.~Currie, A.~Gehrmann-De~Ridder, T.~Gehrmann, E.~W.~N. Glover, A.~Huss, and
  J.~a. Pires, {\it {Infrared sensitivity of single jet inclusive production at
  hadron colliders}},  {\em JHEP} {\bf 10} (2018) 155,
  [\href{http://arxiv.org/abs/1807.03692}{{\tt arXiv:1807.03692}}].

\bibitem{Czakon:2021ohs}
M.~L. Czakon, T.~Generet, A.~Mitov, and R.~Poncelet, {\it {B-hadron
  hadro-production in NNLO QCD: application to LHC $t\bar{t}$ events with
  leptonic decays}},  \href{http://arxiv.org/abs/2102.08267}{{\tt
  arXiv:2102.08267}}.

\bibitem{Czakon:2019tmo}
M.~Czakon, A.~van Hameren, A.~Mitov, and R.~Poncelet, {\it {Single-jet
  inclusive rates with exact color at $ \mathcal{O} $ ($ {\alpha}_s^4 $)}},
  {\em JHEP} {\bf 10} (2019) 262, [\href{http://arxiv.org/abs/1907.12911}{{\tt
  arXiv:1907.12911}}].

\bibitem{Goyal:2024tmo}
S.~Goyal, R.~N. Lee, S.-O. Moch, V.~Pathak, N.~Rana, and V.~Ravindran, {\it
  {NNLO QCD corrections to polarized semi-inclusive DIS}},
  \href{http://arxiv.org/abs/2404.09959}{{\tt arXiv:2404.09959}}.

\bibitem{Goyal:2023zdi}
S.~Goyal, S.-O. Moch, V.~Pathak, N.~Rana, and V.~Ravindran, {\it
  {Next-to-Next-to-Leading Order QCD Corrections to Semi-Inclusive
  Deep-Inelastic Scattering}},  {\em Phys. Rev. Lett.} {\bf 132} (2024), no.~25
  251902, [\href{http://arxiv.org/abs/2312.17711}{{\tt arXiv:2312.17711}}].

\bibitem{Bonino:2024wgg}
L.~Bonino, T.~Gehrmann, M.~L\"ochner, K.~Sch\"onwald, and G.~Stagnitto, {\it
  {Polarized semi-inclusive deep-inelastic scattering at NNLO in QCD}},
  \href{http://arxiv.org/abs/2404.08597}{{\tt arXiv:2404.08597}}.

\bibitem{Bonino:2024qbh}
L.~Bonino, T.~Gehrmann, and G.~Stagnitto, {\it {Semi-Inclusive Deep-Inelastic
  Scattering at Next-to-Next-to-Leading Order in QCD}},  {\em Phys. Rev. Lett.}
  {\bf 132} (2024), no.~25 251901, [\href{http://arxiv.org/abs/2401.16281}{{\tt
  arXiv:2401.16281}}].

\bibitem{Hannesdottir:2022rsl}
H.~S. Hannesdottir, A.~Pathak, M.~D. Schwartz, and I.~W. Stewart, {\it
  {Prospects for strong coupling measurement at hadron colliders using
  soft-drop jet mass}},  {\em JHEP} {\bf 04} (2023) 087,
  [\href{http://arxiv.org/abs/2210.04901}{{\tt arXiv:2210.04901}}].

\bibitem{Lee:2023xzv}
K.~Lee, I.~Moult, F.~Ringer, and W.~J. Waalewijn, {\it {A formalism for
  extracting track functions from jet measurements}},  {\em JHEP} {\bf 01}
  (2024) 194, [\href{http://arxiv.org/abs/2308.00028}{{\tt arXiv:2308.00028}}].

\bibitem{Craft:2022kdo}
E.~Craft, K.~Lee, B.~Me\c{c}aj, and I.~Moult, {\it {Beautiful and Charming
  Energy Correlators}},  \href{http://arxiv.org/abs/2210.09311}{{\tt
  arXiv:2210.09311}}.

\bibitem{Chien:2018lmv}
Y.-T. Chien, D.~Kang, K.~Lee, and Y.~Makris, {\it {Subtracted Cumulants:
  Mitigating Large Background in Jet Substructure}},  {\em Phys. Rev. D} {\bf
  100} (2019), no.~7 074030, [\href{http://arxiv.org/abs/1812.06977}{{\tt
  arXiv:1812.06977}}].

\bibitem{Lee:2022ige}
K.~Lee, B.~Me\c{c}aj, and I.~Moult, {\it {Conformal Colliders Meet the LHC}},
  \href{http://arxiv.org/abs/2205.03414}{{\tt arXiv:2205.03414}}.

\bibitem{Lee:2023npz}
K.~Lee and I.~Moult, {\it {Energy Correlators Taking Charge}},
  \href{http://arxiv.org/abs/2308.00746}{{\tt arXiv:2308.00746}}.

\bibitem{Aschenauer:2019uex}
E.-C. Aschenauer, K.~Lee, B.~S. Page, and F.~Ringer, {\it {Jet angularities in
  photoproduction at the Electron-Ion Collider}},  {\em Phys. Rev. D} {\bf 101}
  (2020), no.~5 054028, [\href{http://arxiv.org/abs/1910.11460}{{\tt
  arXiv:1910.11460}}].

\bibitem{Kang:2018qra}
Z.-B. Kang, K.~Lee, and F.~Ringer, {\it {Jet angularity measurements for single
  inclusive jet production}},  {\em JHEP} {\bf 04} (2018) 110,
  [\href{http://arxiv.org/abs/1801.00790}{{\tt arXiv:1801.00790}}].

\bibitem{Cal:2020flh}
P.~Cal, K.~Lee, F.~Ringer, and W.~J. Waalewijn, {\it {Jet energy drop}},  {\em
  JHEP} {\bf 11} (2020) 012, [\href{http://arxiv.org/abs/2007.12187}{{\tt
  arXiv:2007.12187}}].

\bibitem{Kang:2020xyq}
Z.-B. Kang, K.~Lee, and F.~Zhao, {\it {Polarized jet fragmentation functions}},
   {\em Phys. Lett. B} {\bf 809} (2020) 135756,
  [\href{http://arxiv.org/abs/2005.02398}{{\tt arXiv:2005.02398}}].

\bibitem{Kang:2018vgn}
Z.-B. Kang, K.~Lee, X.~Liu, and F.~Ringer, {\it {Soft drop groomed jet
  angularities at the LHC}},  {\em Phys. Lett.} {\bf B793} (2019) 41--47,
  [\href{http://arxiv.org/abs/1811.06983}{{\tt arXiv:1811.06983}}].

\bibitem{Kang:2018jwa}
Z.-B. Kang, K.~Lee, X.~Liu, and F.~Ringer, {\it {The groomed and ungroomed jet
  mass distribution for inclusive jet production at the LHC}},  {\em JHEP} {\bf
  10} (2018) 137, [\href{http://arxiv.org/abs/1803.03645}{{\tt
  arXiv:1803.03645}}].

\bibitem{Kang:2019prh}
Z.-B. Kang, K.~Lee, X.~Liu, D.~Neill, and F.~Ringer, {\it {The soft drop
  groomed jet radius at NLL}},  \href{http://arxiv.org/abs/1908.01783}{{\tt
  arXiv:1908.01783}}.

\bibitem{Cal:2021fla}
P.~Cal, K.~Lee, F.~Ringer, and W.~J. Waalewijn, {\it {The soft drop momentum
  sharing fraction $z_g$ beyond leading-logarithmic accuracy}},
  \href{http://arxiv.org/abs/2106.04589}{{\tt arXiv:2106.04589}}.

\bibitem{Cal:2019gxa}
P.~Cal, D.~Neill, F.~Ringer, and W.~J. Waalewijn, {\it {Calculating the angle
  between jet axes}},  {\em JHEP} {\bf 04} (2020) 211,
  [\href{http://arxiv.org/abs/1911.06840}{{\tt arXiv:1911.06840}}].

\bibitem{Kang:2016ehg}
Z.-B. Kang, F.~Ringer, and I.~Vitev, {\it {Jet substructure using
  semi-inclusive jet functions in SCET}},  {\em JHEP} {\bf 11} (2016) 155,
  [\href{http://arxiv.org/abs/1606.07063}{{\tt arXiv:1606.07063}}].

\bibitem{Cal:2019hjc}
P.~Cal, F.~Ringer, and W.~J. Waalewijn, {\it {The jet shape at NLL'}},  {\em
  JHEP} {\bf 05} (2019) 143, [\href{http://arxiv.org/abs/1901.06389}{{\tt
  arXiv:1901.06389}}].

\bibitem{Kang:2017mda}
Z.-B. Kang, F.~Ringer, and W.~J. Waalewijn, {\it {The Energy Distribution of
  Subjets and the Jet Shape}},  {\em JHEP} {\bf 07} (2017) 064,
  [\href{http://arxiv.org/abs/1705.05375}{{\tt arXiv:1705.05375}}].

\bibitem{Mehtar-Tani:2024smp}
Y.~Mehtar-Tani, F.~Ringer, B.~Singh, and V.~Vaidya, {\it {Factorization for jet
  production in heavy-ion collisions}},
  \href{http://arxiv.org/abs/2409.05957}{{\tt arXiv:2409.05957}}.

\bibitem{ALICE:2019qyj}
{\bf ALICE} Collaboration, S.~Acharya et~al., {\it {Measurements of inclusive
  jet spectra in pp and central Pb-Pb collisions at $\sqrt{s_{\rm{NN}}}$ = 5.02
  TeV}},  {\em Phys. Rev. C} {\bf 101} (2020), no.~3 034911,
  [\href{http://arxiv.org/abs/1909.09718}{{\tt arXiv:1909.09718}}].

\bibitem{CMS:2020caw}
{\bf CMS} Collaboration, A.~M. Sirunyan et~al., {\it {Dependence of inclusive
  jet production on the anti-k$_{T}$ distance parameter in pp collisions at $
  \sqrt{\mathrm{s}} $ = 13 TeV}},  {\em JHEP} {\bf 12} (2020) 082,
  [\href{http://arxiv.org/abs/2005.05159}{{\tt arXiv:2005.05159}}].

\bibitem{ALICE:2023waz}
{\bf ALICE} Collaboration, S.~Acharya et~al., {\it {Measurement of the radius
  dependence of charged-particle jet suppression in Pb\textendash{}Pb
  collisions at sNN=5.02TeV}},  {\em Phys. Lett. B} {\bf 849} (2024) 138412,
  [\href{http://arxiv.org/abs/2303.00592}{{\tt arXiv:2303.00592}}].

\bibitem{CMS:2021vui}
{\bf CMS} Collaboration, A.~M. Sirunyan et~al., {\it {First measurement of
  large area jet transverse momentum spectra in heavy-ion collisions}},  {\em
  JHEP} {\bf 05} (2021) 284, [\href{http://arxiv.org/abs/2102.13080}{{\tt
  arXiv:2102.13080}}].

\bibitem{Collins:1981ta}
J.~C. Collins and G.~F. Sterman, {\it {Soft Partons in {QCD}}},  {\em Nucl.
  Phys. B} {\bf 185} (1981) 172--188.

\bibitem{Bodwin:1984hc}
G.~T. Bodwin, {\it {Factorization of the Drell-Yan Cross-Section in
  Perturbation Theory}},  {\em Phys. Rev. D} {\bf 31} (1985) 2616. [Erratum:
  Phys.Rev.D 34, 3932 (1986)].

\bibitem{Collins:1985ue}
J.~C. Collins, D.~E. Soper, and G.~F. Sterman, {\it {Factorization for Short
  Distance Hadron - Hadron Scattering}},  {\em Nucl. Phys. B} {\bf 261} (1985)
  104--142.

\bibitem{Collins:1988ig}
J.~C. Collins, D.~E. Soper, and G.~F. Sterman, {\it {Soft Gluons and
  Factorization}},  {\em Nucl. Phys. B} {\bf 308} (1988) 833--856.

\bibitem{Collins:1989gx}
J.~C. Collins, D.~E. Soper, and G.~F. Sterman, {\it {Factorization of Hard
  Processes in QCD}},  {\em Adv. Ser. Direct. High Energy Phys.} {\bf 5} (1989)
  1--91, [\href{http://arxiv.org/abs/hep-ph/0409313}{{\tt hep-ph/0409313}}].

\bibitem{Collins:2011zzd}
J.~Collins, {\em {Foundations of perturbative QCD}}, vol.~32.
\newblock Cambridge University Press, 11, 2013.

\bibitem{Nayak:2005rt}
G.~C. Nayak, J.-W. Qiu, and G.~F. Sterman, {\it {Fragmentation, NRQCD and NNLO
  factorization analysis in heavy quarkonium production}},  {\em Phys. Rev. D}
  {\bf 72} (2005) 114012, [\href{http://arxiv.org/abs/hep-ph/0509021}{{\tt
  hep-ph/0509021}}].

\bibitem{Dasgupta:2014yra}
M.~Dasgupta, F.~Dreyer, G.~P. Salam, and G.~Soyez, {\it {Small-radius jets to
  all orders in QCD}},  {\em JHEP} {\bf 04} (2015) 039,
  [\href{http://arxiv.org/abs/1411.5182}{{\tt arXiv:1411.5182}}].

\bibitem{Dasgupta:2016bnd}
M.~Dasgupta, F.~A. Dreyer, G.~P. Salam, and G.~Soyez, {\it {Inclusive jet
  spectrum for small-radius jets}},  {\em JHEP} {\bf 06} (2016) 057,
  [\href{http://arxiv.org/abs/1602.01110}{{\tt arXiv:1602.01110}}].

\bibitem{Kang:2016mcy}
Z.-B. Kang, F.~Ringer, and I.~Vitev, {\it {The semi-inclusive jet function in
  SCET and small radius resummation for inclusive jet production}},  {\em JHEP}
  {\bf 10} (2016) 125, [\href{http://arxiv.org/abs/1606.06732}{{\tt
  arXiv:1606.06732}}].

\bibitem{Liu:2018ktv}
X.~Liu, S.-O. Moch, and F.~Ringer, {\it {Phenomenology of single-inclusive jet
  production with jet radius and threshold resummation}},  {\em Phys. Rev. D}
  {\bf 97} (2018), no.~5 056026, [\href{http://arxiv.org/abs/1801.07284}{{\tt
  arXiv:1801.07284}}].

\bibitem{vanBeekveld:2024jnx}
M.~van Beekveld, M.~Dasgupta, B.~K. El-Menoufi, J.~Helliwell, A.~Karlberg, and
  P.~F. Monni, {\it {Two-loop anomalous dimensions for small-$R$ jet versus
  hadronic fragmentation functions}},
  \href{http://arxiv.org/abs/2402.05170}{{\tt arXiv:2402.05170}}.

\bibitem{vanBeekveld:2024qxs}
M.~van Beekveld, M.~Dasgupta, B.~K. El-Menoufi, J.~Helliwell, P.~F. Monni, and
  G.~P. Salam, {\it {A collinear shower algorithm for NSL non-singlet
  fragmentation}},  \href{http://arxiv.org/abs/2409.08316}{{\tt
  arXiv:2409.08316}}.

\bibitem{Bauer:2000ew}
C.~W. Bauer, S.~Fleming, and M.~E. Luke, {\it {Summing Sudakov logarithms in B
  ---\ensuremath{>} X(s gamma) in effective field theory}},  {\em Phys. Rev. D}
  {\bf 63} (2000) 014006, [\href{http://arxiv.org/abs/hep-ph/0005275}{{\tt
  hep-ph/0005275}}].

\bibitem{Bauer:2000yr}
C.~W. Bauer, S.~Fleming, D.~Pirjol, and I.~W. Stewart, {\it {An Effective field
  theory for collinear and soft gluons: Heavy to light decays}},  {\em Phys.
  Rev. D} {\bf 63} (2001) 114020,
  [\href{http://arxiv.org/abs/hep-ph/0011336}{{\tt hep-ph/0011336}}].

\bibitem{Bauer:2001ct}
C.~W. Bauer and I.~W. Stewart, {\it {Invariant operators in collinear effective
  theory}},  {\em Phys. Lett.} {\bf B516} (2001) 134--142,
  [\href{http://arxiv.org/abs/hep-ph/0107001}{{\tt hep-ph/0107001}}].

\bibitem{Bauer:2001yt}
C.~W. Bauer, D.~Pirjol, and I.~W. Stewart, {\it {Soft collinear factorization
  in effective field theory}},  {\em Phys. Rev. D} {\bf 65} (2002) 054022,
  [\href{http://arxiv.org/abs/hep-ph/0109045}{{\tt hep-ph/0109045}}].

\bibitem{Rothstein:2016bsq}
I.~Z. Rothstein and I.~W. Stewart, {\it {An Effective Field Theory for Forward
  Scattering and Factorization Violation}},  {\em JHEP} {\bf 08} (2016) 025,
  [\href{http://arxiv.org/abs/1601.04695}{{\tt arXiv:1601.04695}}].

\bibitem{Lee:2024icn}
K.~Lee, I.~Moult, and X.~Zhang, {\it {Revisiting Single Inclusive Jet
  Production: Timelike Factorization and Reciprocity}},
  \href{http://arxiv.org/abs/2409.19045}{{\tt arXiv:2409.19045}}.

\bibitem{Dixon:2019uzg}
L.~J. Dixon, I.~Moult, and H.~X. Zhu, {\it {Collinear limit of the
  energy-energy correlator}},  {\em Phys. Rev. D} {\bf 100} (2019), no.~1
  014009, [\href{http://arxiv.org/abs/1905.01310}{{\tt arXiv:1905.01310}}].

\bibitem{Chen:2020vvp}
H.~Chen, I.~Moult, X.~Zhang, and H.~X. Zhu, {\it {Rethinking jets with energy
  correlators: Tracks, resummation, and analytic continuation}},  {\em Phys.
  Rev. D} {\bf 102} (2020), no.~5 054012,
  [\href{http://arxiv.org/abs/2004.11381}{{\tt arXiv:2004.11381}}].

\bibitem{Belitsky:2013ofa}
A.~Belitsky, S.~Hohenegger, G.~Korchemsky, E.~Sokatchev, and A.~Zhiboedov, {\it
  {Energy-Energy Correlations in N=4 Supersymmetric Yang-Mills Theory}},  {\em
  Phys. Rev. Lett.} {\bf 112} (2014), no.~7 071601,
  [\href{http://arxiv.org/abs/1311.6800}{{\tt arXiv:1311.6800}}].

\bibitem{Dixon:2018qgp}
L.~J. Dixon, M.-X. Luo, V.~Shtabovenko, T.-Z. Yang, and H.~X. Zhu, {\it
  {Analytical Computation of Energy-Energy Correlation at Next-to-Leading Order
  in QCD}},  {\em Phys. Rev. Lett.} {\bf 120} (2018), no.~10 102001,
  [\href{http://arxiv.org/abs/1801.03219}{{\tt arXiv:1801.03219}}].

\bibitem{Luo:2019nig}
M.-X. Luo, V.~Shtabovenko, T.-Z. Yang, and H.~X. Zhu, {\it {Analytic
  Next-To-Leading Order Calculation of Energy-Energy Correlation in
  Gluon-Initiated Higgs Decays}},  {\em JHEP} {\bf 06} (2019) 037,
  [\href{http://arxiv.org/abs/1903.07277}{{\tt arXiv:1903.07277}}].

\bibitem{Henn:2019gkr}
J.~Henn, E.~Sokatchev, K.~Yan, and A.~Zhiboedov, {\it {Energy-energy
  correlation in $N$=4 super Yang-Mills theory at next-to-next-to-leading
  order}},  {\em Phys. Rev. D} {\bf 100} (2019), no.~3 036010,
  [\href{http://arxiv.org/abs/1903.05314}{{\tt arXiv:1903.05314}}].

\bibitem{Neill:2021std}
D.~Neill, F.~Ringer, and N.~Sato, {\it {Leading jets and energy loss}},  {\em
  JHEP} {\bf 07} (2021) 041, [\href{http://arxiv.org/abs/2103.16573}{{\tt
  arXiv:2103.16573}}].

\bibitem{Chen:2021uws}
Y.~Chen et~al., {\it {Jet energy spectrum and substructure in e+e- collisions
  at 91.2 GeV with ALEPH Archived Data}},  {\em JHEP} {\bf 06} (2022) 008,
  [\href{http://arxiv.org/abs/2111.09914}{{\tt arXiv:2111.09914}}].

\bibitem{Gribov:1972ri}
V.~N. Gribov and L.~N. Lipatov, {\it {Deep inelastic e p scattering in
  perturbation theory}},  {\em Sov. J. Nucl. Phys.} {\bf 15} (1972) 438--450.

\bibitem{Dokshitzer:1977sg}
Y.~L. Dokshitzer, {\it {Calculation of the Structure Functions for Deep
  Inelastic Scattering and e+ e- Annihilation by Perturbation Theory in Quantum
  Chromodynamics.}},  {\em Sov. Phys. JETP} {\bf 46} (1977) 641--653.

\bibitem{Altarelli:1977zs}
G.~Altarelli and G.~Parisi, {\it {Asymptotic Freedom in Parton Language}},
  {\em Nucl. Phys. B} {\bf 126} (1977) 298--318.

\bibitem{Mitov:2006ic}
A.~Mitov, S.~Moch, and A.~Vogt, {\it {Next-to-Next-to-Leading Order Evolution
  of Non-Singlet Fragmentation Functions}},  {\em Phys. Lett.} {\bf B638}
  (2006) 61--67, [\href{http://arxiv.org/abs/hep-ph/0604053}{{\tt
  hep-ph/0604053}}].

\bibitem{Mitov:2006wy}
A.~Mitov and S.~Moch, {\it {QCD Corrections to Semi-Inclusive Hadron Production
  in Electron-Positron Annihilation at Two Loops}},  {\em Nucl. Phys.} {\bf
  B751} (2006) 18--52, [\href{http://arxiv.org/abs/hep-ph/0604160}{{\tt
  hep-ph/0604160}}].

\bibitem{Chen:2020uvt}
H.~Chen, T.-Z. Yang, H.~X. Zhu, and Y.~J. Zhu, {\it {Analytic Continuation and
  Reciprocity Relation for Collinear Splitting in QCD}},
  \href{http://arxiv.org/abs/2006.10534}{{\tt arXiv:2006.10534}}.

\bibitem{Moch:2007tx}
S.~Moch and A.~Vogt, {\it {On third-order timelike splitting functions and
  top-mediated Higgs decay into hadrons}},  {\em Phys. Lett. B} {\bf 659}
  (2008) 290--296, [\href{http://arxiv.org/abs/0709.3899}{{\tt
  arXiv:0709.3899}}].

\bibitem{Almasy:2011eq}
A.~A. Almasy, S.~Moch, and A.~Vogt, {\it {On the Next-to-Next-to-Leading Order
  Evolution of Flavour-Singlet Fragmentation Functions}},  {\em Nucl. Phys. B}
  {\bf 854} (2012) 133--152, [\href{http://arxiv.org/abs/1107.2263}{{\tt
  arXiv:1107.2263}}].

\bibitem{Salam:2008qg}
G.~P. Salam and J.~Rojo, {\it {A Higher Order Perturbative Parton Evolution
  Toolkit (HOPPET)}},  {\em Comput. Phys. Commun.} {\bf 180} (2009) 120--156,
  [\href{http://arxiv.org/abs/0804.3755}{{\tt arXiv:0804.3755}}].

\bibitem{Botje:2010ay}
M.~Botje, {\it {QCDNUM: Fast QCD Evolution and Convolution}},  {\em Comput.
  Phys. Commun.} {\bf 182} (2011) 490--532,
  [\href{http://arxiv.org/abs/1005.1481}{{\tt arXiv:1005.1481}}].

\bibitem{Bertone:2013vaa}
V.~Bertone, S.~Carrazza, and J.~Rojo, {\it {APFEL: A PDF Evolution Library with
  QED corrections}},  {\em Comput. Phys. Commun.} {\bf 185} (2014) 1647--1668,
  [\href{http://arxiv.org/abs/1310.1394}{{\tt arXiv:1310.1394}}].

\bibitem{Bolzoni:2013rsa}
P.~Bolzoni, B.~A. Kniehl, and A.~V. Kotikov, {\it {Average gluon and quark jet
  multiplicities at higher orders}},  {\em Nucl. Phys. B} {\bf 875} (2013)
  18--44, [\href{http://arxiv.org/abs/1305.6017}{{\tt arXiv:1305.6017}}].

\bibitem{Bolzoni:2012ii}
P.~Bolzoni, B.~A. Kniehl, and A.~V. Kotikov, {\it {Gluon and quark jet
  multiplicities at N$^3$LO+NNLL}},  {\em Phys. Rev. Lett.} {\bf 109} (2012)
  242002, [\href{http://arxiv.org/abs/1209.5914}{{\tt arXiv:1209.5914}}].

\bibitem{Chen:2023zlx}
W.~Chen, J.~Gao, Y.~Li, Z.~Xu, X.~Zhang, and H.~X. Zhu, {\it {NNLL Resummation
  for Projected Three-Point Energy Correlator}},
  \href{http://arxiv.org/abs/2307.07510}{{\tt arXiv:2307.07510}}.

\bibitem{Jaarsma:2023ell}
M.~Jaarsma, Y.~Li, I.~Moult, W.~J. Waalewijn, and H.~X. Zhu, {\it {Energy
  correlators on tracks: resummation and non-perturbative effects}},  {\em
  JHEP} {\bf 12} (2023) 087, [\href{http://arxiv.org/abs/2307.15739}{{\tt
  arXiv:2307.15739}}].

\bibitem{Remiddi:1999ew}
E.~Remiddi and J.~A.~M. Vermaseren, {\it {Harmonic polylogarithms}},  {\em Int.
  J. Mod. Phys. A} {\bf 15} (2000) 725--754,
  [\href{http://arxiv.org/abs/hep-ph/9905237}{{\tt hep-ph/9905237}}].

\bibitem{Vollinga:2004sn}
J.~Vollinga and S.~Weinzierl, {\it {Numerical evaluation of multiple
  polylogarithms}},  {\em Comput. Phys. Commun.} {\bf 167} (2005) 177,
  [\href{http://arxiv.org/abs/hep-ph/0410259}{{\tt hep-ph/0410259}}].

\bibitem{Gehrmann:2001pz}
T.~Gehrmann and E.~Remiddi, {\it {Numerical evaluation of harmonic
  polylogarithms}},  {\em Comput. Phys. Commun.} {\bf 141} (2001) 296--312,
  [\href{http://arxiv.org/abs/hep-ph/0107173}{{\tt hep-ph/0107173}}].

\bibitem{Bauer:2000cp}
C.~W. Bauer, A.~Frink, and R.~Kreckel, {\it {Introduction to the GiNaC
  framework for symbolic computation within the C++ programming language}},
  {\em J. Symb. Comput.} {\bf 33} (2002) 1--12,
  [\href{http://arxiv.org/abs/cs/0004015}{{\tt cs/0004015}}].

\bibitem{Dasgupta:2007wa}
M.~Dasgupta, L.~Magnea, and G.~P. Salam, {\it {Non-perturbative QCD effects in
  jets at hadron colliders}},  {\em JHEP} {\bf 02} (2008) 055,
  [\href{http://arxiv.org/abs/0712.3014}{{\tt arXiv:0712.3014}}].

\bibitem{Lee:2024esz}
K.~Lee, A.~Pathak, I.~Stewart, and Z.~Sun, {\it {Nonperturbative Effects in
  Energy Correlators: From Characterizing Confinement Transition to Improving
  $\alpha_s$ Extraction}},  \href{http://arxiv.org/abs/2405.19396}{{\tt
  arXiv:2405.19396}}.

\bibitem{Maitre:2005uu}
D.~Maitre, {\it {HPL, a mathematica implementation of the harmonic
  polylogarithms}},  {\em Comput. Phys. Commun.} {\bf 174} (2006) 222--240,
  [\href{http://arxiv.org/abs/hep-ph/0507152}{{\tt hep-ph/0507152}}].

\bibitem{Ablinger:2009ovq}
J.~Ablinger, {\it {A Computer Algebra Toolbox for Harmonic Sums Related to
  Particle Physics}},  Master's thesis, Linz U., 2009.

\bibitem{Hoschele:2013pvt}
M.~H\"oschele, J.~Hoff, A.~Pak, M.~Steinhauser, and T.~Ueda, {\it {MT: A
  Mathematica package to compute convolutions}},  {\em Comput. Phys. Commun.}
  {\bf 185} (2014) 528--539, [\href{http://arxiv.org/abs/1307.6925}{{\tt
  arXiv:1307.6925}}].

\end{thebibliography}\endgroup
\bibliographystyle{JHEP}

\end{document}